\documentstyle[emulateapj]{article}

\received{}
\revised{}
\accepted{}
\journalid{}{}
\articleid{}{}
\paperid{}
\cpright{}{}
\ccc{}

\slugcomment{Submitted to the Astrophysical Journal}

\lefthead{Wood et al.}
\righthead{Infrared Signatures of Protoplanetary Disk Evolution}

\begin{document}

\title{Infrared Signatures of Protoplanetary Disk Evolution}

\author{Kenneth Wood\altaffilmark{1,2}, C.J.~Lada\altaffilmark{2}, 
J.E.~Bjorkman\altaffilmark{3}, Scott~J.~Kenyon\altaffilmark{2}, 
Barbara Whitney\altaffilmark{4}, Michael~J.~Wolff\altaffilmark{4}  }

\altaffiltext{1}{School of Physics \& Astronomy, University of St Andrews,
North Haugh, St Andrews, Kingdom of Fife, KY16 9SS, Scotland;
kw25@st-andrews.ac.uk}

\altaffiltext{2}{Harvard-Smithsonian Center for Astrophysics, 
60 Garden Street, Cambridge, MA~02138; 
clada@cfa.harvard.edu; kenyon@payne.harvard.edu}

\altaffiltext{3}{Ritter Observatory, Department of Physics \& Astronomy, 
University of Toledo, Toledo, OH 43606; jon@astro.utoledo.edu}

\altaffiltext{4}{Space Science Institute, 3100 Marine Street, Suite A353, 
Boulder, CO~80303; bwhitney@colorado.edu, wolff@colorado.edu}

\authoremail{kw25@st-andrews.ac.uk}

\begin{abstract}

We investigate the observational signatures of a straightforward evolutionary 
scenario for protoplanetary disks in which the disk mass of small 
($\la 50\mu{\rm m}$) particles decreases homologously with time, 
but the disk structure and stellar parameters do not change.  
Our goal is to identify optimal infrared spectral indicators of the 
existence of disks, their structure, and mass evolution that may be tested 
with the upcoming {\it SIRTF} mission.  
We present simulated spectral energy distributions and colors over a wide 
range of masses, $10^{-8}M_\odot \le M_{\rm disk} \le 10^{-1}M_\odot$.  
Our Monte Carlo radiative equilibrium techniques enable us to explore 
the wide range of optical depths of these disks 
and incorporate multiple, anisotropic dust scattering.  
The SED is most sensitive to disk mass in the far-IR and longer wavelengths, 
which is already known from millimeter and radio observations.  
As the disk mass decreases, the excess emission of the 
disk over the stellar photosphere diminishes more 
rapidly at the longest rather than at short wavelengths.  
At near-infrared wavelengths, the disk 
remains optically thick to stellar radiation over a wide range of disk 
mass, resulting in a slower decline in the SED in this spectral regime.  
Therefore, near-IR excesses ($K-L$) provide a robust means of 
detecting disks in star clusters down to $M_{\rm disk}\sim 10^{-7}M_\odot$, 
while the far-IR excess probes the disk mass, the 
caveat being that large inner disk holes can decrease the near-IR disk 
emission.   

Varying other disk 
parameters (outer radius, flaring, dust size distribution) alter the SED 
quantitatively, but do 
not change our general conclusions on the evolution of SEDs and colors with 
the mass of small particles in the disk.  
Reducing the disk mass results in a clear progression in color-color 
diagrams with low mass disks displaying the bluest colors.  We interpret 
color-color diagrams for Taurus-Auriga sources in the context of decreasing 
disk mass.  
Different viewing angles yield degeneracies in the color-mass relationship, 
but highly inclined disks are very faint and red and are readily 
identified in color-magnitude diagrams.  

\end{abstract}

\keywords{radiative transfer --- scattering --- 
accretion, accretion disks --- ISM: dust, extinction --- 
stars: pre-main-sequence}

\section{Introduction}

Protoplanetary disks are primarily detected through their signature infrared 
excess emission relative to a stellar photosphere (e.g., Mendoza 1968; 
Rydgren, Strom, \& Strom 1976; Cohen \& Kuhi 1979; Rucinski 1985).  The IR 
excesses arise from the reprocessing of stellar radiation and the liberation 
of accretion luminosity within the disk (e.g., Lynden-Bell \& Pringle 1974; 
Adams \& Shu 1986; 
Adams, Lada, \& Shu 1987; Kenyon \& Hartmann 1987; Adams, Emerson, \& Fuller 
1990; Lada \& Adams 1992; Hillenbrand et al. 1992).  
At other wavelengths, disk masses and velocity structures have been measured 
with radio and mm interferometers (Beckwith et al. 1990; 
Beckwith \& Sargent 1993; Koerner, 
Sargent, \& Beckwith 1993; Koerner \& Sargent 1995; Dutery et al. 1996; 
Wilner \& Lay 2000) and more recently, high 
resolution {\it HST} observations and ground based adaptive optics and 
speckle imaging techniques have imaged disks via scattered light 
(O'Dell, Wen, \& Hu 1993; Burrows et al. 1996; Stapelfeldt et al. 1998; 
Padgett et al. 1999; Krist et al. 2000; Cotera et al. 2001; 
Grady et al. 2000; Roddier et al. 1996; Koresko 1998).

The disks that are detected range in mass, structure, and formation 
mechanisms.  Massive, optically thick, flaring circumstellar disks, 
$M_{\rm disk} \ga 10^{-3}M_\odot$, are a natural product 
of the star formation process (e.g., Shu, Adams, \& Lizano 1987) and 
are detected around pre-main sequence Classical T-Tauri stars.  
Low mass ``debris disks'' ($M_{\rm disk} \la 10^{-5}M_\odot$) 
are detected around older main sequence stars (Backman \& Paresce 1993).  
The dust that gives rise to the IR excesses and scattered 
light images of debris disks is believed to form from collisions of 
planetessimals within the disks.  

The advent of large IR detector arrays now allows for simultaneous 
multiwavelength observations of large numbers of disks in star clusters 
(e.g., Haisch, Lada, \& Lada 2000).  
The upcoming {\it SIRTF} mission will 
greatly increase the wavelength coverage for studying clusters, thereby 
enabling detailed studies of disks over a wide range in age and mass.  
We therefore need to know what spectral indicators are best suited for 
detecting disks and determining their mass evolution.

In this paper our working 
hypothesis is that through time the disk mass decreases but the disk 
structure and stellar luminosity do not change.  In our models the dominant 
opacity source for $\lambda\la 100\mu$m is small ($\la 50\mu{\rm m}$) 
grains, so another 
way of stating our model is that we assume the small particle mass traces 
the total mass of gas + dust.  The small grain mass may decrease by the 
accretion of disk material or 
coagulation and growth of large grains and rocks while the gas is either 
depleted or accreted onto the star keeping the gas/dust ratio constant.  
The recent detection of $H_2$ in 
$\beta$ Pictoris' disk and the derived gas/dust $\sim 100$ is consistent 
with our assumption (Thi et al. 2001).

We investigate the observational signatures 
of this simple model for disk evolution and present radiative 
equilibrium models for a range of disk 
masses and viewing angles.  Model spectra are compared with 
semi-analytic approximations for optically thick 
flat and flared disks and predictions are made 
as to what spectral regions are most sensitive to the evolution of disk mass 
and at what wavelengths disk detections are robust.  
We perform simulations for a fiducial disk structure and 
dust size distribution that fits 
the SED of the edge-on disk system HH30~IRS. (Wood et al. 2001).  
\S~2 describes the ingredients of our models, \S~3 presents our 
SED simulations and shows how {\it SIRTF} colors 
are sensitive to disk mass.  \S~4 presents models that explore 
variations in parameters other than disk mass and discusses 
alternative disk evolution models.  \S~5 compares our 
models with color-color diagrams of Taurus-Auriga sources and we 
summarize our findings in \S~6.

\section{Models}

\subsection{Radiative Equilibrium Calculation}

Theoretical temperature distributions and emergent spectra have been 
calculated for T~Tauri disks using a variety of techniques.  
These include approximations for opaque flat disks 
(Adams \& Shu 1986; Adams, Lada, \& Shu 1987; Lada \& Adams 1992) and 
extensions of 
this approach to flared disks (Kenyon \& Hartmann 1987; Chiang \& Goldreich 
1997, 1999); two dimensional radiation transfer techniques 
(Efstathiou \& Rowan-Robinson 1991); 
diffusion approximations (e.g., Sonnhalter, Preibisch, \& 
Yorke 1995; Boss \& Yorke 1996);  
disks which are modeled as ``spherical sectors'' 
(Men'shchikov \& Henning 1997); vertical structure calculations 
performed by dividing the disk into plane parallel annuli (Calvet et al. 
1992; Bell et al. 1997; Bell 1999; 
D'Alessio et al. 1998, 1999); and Monte Carlo techniques (Wolf, 
Henning, \& Stecklum 1999).  Common approximations in many of the 
``traditional'' codes are to conduct the radiation 
transfer in only one direction in the disk and assume the dust 
grains scatter radiation isotropically.  Also many techniques are limited 
to the study of optically thick and hence massive disks.  

Monte Carlo techniques are straightforward to adapt to any geometry, 
mass, and can accurately include polarization and multiple, anisotropic 
scattering.  We use the Monte Carlo radiative equilibrium technique of 
Bjorkman \& Wood (2001) which conserves energy exactly.  For very optically 
thick disks we use our Monte Carlo technique for the upper layers of the 
disk (i.e., the disk ``atmosphere'') and a diffusion approximation for the 
densest midplane regions.  For the most massive disks we simulate, typically 
less than 1\% of the flux requires a diffusion treatment.  More details of 
our adaptation of the Bjorkman \& Wood (2001) technique for simulating 
T~Tauri disks will be presented in Bjorkman, Wood, \& Whitney (2001, in 
preparation).  
An advantage of using Monte Carlo techniques for studying a range of disk 
masses, is that we are not 
restricted to one-dimensional radiation transfer and can therefore 
simulate SEDs of low mass optically thin disks in which radial transport 
of photons is important. The output of 
our code is the disk temperature structure (due to heating by stellar 
photons and accretion luminosity) and the emergent SED and polarization 
spectrum at a range of viewing angles.  A calculation of the 
hydrostatic disk structure (e.g., Chiang \& Goldreich 1997; 
D'Alessio et al. 1999) 
can be included in the Monte Carlo technique, but this would require 
an iterative scheme.  At present we have not implemented such a scheme and 
instead perform the radiative equilibrium calculation for a fixed 
disk geometry.

\subsection{Disk Structure}

The determination of disk structure from fitting 
SEDs and scattered light images does not yield a single structure and 
dust size distribution that applies to all disks.  Some systems are 
fit with passive flat disks (e.g., Adams et al. 1987; Adams et al. 1990; 
Miyake \& Nakagwa 
1995), while others require flared disks heated by starlight and 
accretion luminosity (e.g., Kenyon \& Hartmann 1987; Burrows et al. 1996; 
Stapelfeldt et al. 1998).  The scattered light image of the edge-on disk of 
HH30~IRS (Burrows et al. 1996) has for the first time allowed the vertical 
structure of a protoplanetary disk to be studied directly.  We therefore 
adopt the HH30~IRS disk as our fiducial model for our SED models.  
A fixed disk density structure which fits the 
scattered light images (Burrows et al. 1996; Cotera et al. 2001) and 
SED (Stapelfeldt \& Monetti 1999; Wood et al. 2001) of HH30~IRS is
\begin{equation}
\rho=\rho_0 \left ({R_\star\over{\varpi}}\right )^{\alpha}
\exp{ -{1\over 2} [z/h(\varpi )]^2  }  
\; ,
\end{equation}
where $\varpi$ is the radial coordinate in the disk midplane and the 
scale height increases with radius,
$h=h_0\left ( {\varpi /{R_\star}} \right )^\beta$.  
For the HH30~IRS disk we adopt $\beta=1.25$, $\alpha=2.25$, 
and $h_0=0.017R_\star$, giving $h(100{\rm AU})=17$AU.

In our simulations the inner edge of the disk is truncated at the dust 
destruction radius, $R_{\rm dust}$.  Assuming $T_\star = 4000$K, 
$R_\star = 2R_\odot$, and circumstellar dust 
sublimates at 1600K, then $R_{\rm dust}\approx 8 R_\star$.  This is 
larger than the dust destruction radius for optically thin dust because the 
reprocessed emission from the disk provides additional heating over 
and above the direct stellar radiation, increasing the size of the dust 
destruction zone (a detailed discussion of the shape of the dust destruction 
region will be presented in Bjorkman et al. 2001, in preparation).  
In currently popular magnetic 
accretion models the disk is truncated at a radius $R_0$, which may not 
be equal to $R_{\rm dust}$.  If $R_0 < R_{\rm dust}$ there 
will be a gas disk extending from $R_0$ to $R_{\rm dust}$ which may give 
rise to additional IR emission.  
Our models therefore assume that any material within 
$R_{\rm dust}$ is optically thin, which is a good approximation for low 
mass disks.  For high mass disks the gas may be optically thick producing 
larger near-IR excesses.

\subsection{Adopted Circumstellar Dust Properties}

Recent modeling of HST images (Cotera et al. 2001) and 
the SED of HH30~IRS (Wood et al. 2001) indicates that 
the circumstellar dust size distribution 
extends to larger grain radii than typical ISM grains.  
This is in agreement with many other observations 
indicating grain growth within protoplanetary 
disks (e.g., Beckwith et al. 1990; Beckwith \& Sargent 1991).  
This paper primarily investigates the effects of disk mass 
on the SED and adopts circumstellar dust properties that 
reproduce the HH30~IRS SED.  The dust model (chemical 
composition, mathematical form for the size distribution, 
calculation of opacity and scattering parameters, etc) 
is described in Wood et al. (2001) and we 
only summarize the main features here.  
Specifically, we adopt, a size distribution
\begin{equation}
n(a) \,\, da  =  C_i \, a^{-p}\,  \exp{(-[a/a_c]^q)} \,\ da\; ,
\end{equation}
with $p=3.5$, $q=0.6$, $a_c=50\mu{\rm m}$, 
$a_{\rm min}=0.01\mu{\rm m}$, and $a_{\rm max}=1$mm.  
The exponential scalelength, $a_c$, yields dust 
particle sizes extending up to and in excess of $50\mu{\rm m}$.  
Figure~1 shows the 
wavelength dependence of the opacity, scattering albedo, 
and Heyney-Greenstein phase function asymmetry parameter 
(Heyney \& Greenstein 1941) for this size distribution. 

Recent {\it Infrared Space Observatory} spectra 
of Herbig Ae/Be stars (e.g., Meuss et al. 2000; Chiang et al. 
2001; van den Ancker et al. 2000) are now allowing the 
circumstellar dust chemistry to be studied.  Because we 
adopt the dust properties of Fig.~1 for our simulations, 
we have not investigated the effects of different chemical 
compositions on the the resulting disk SEDs.  How the 
circumstellar chemistry effects the SED is an interesting 
problem, but it is beyond the scope of this paper and we 
present models for different disk masses 
and the dust properties shown in Figure~1.  
In addition, our models do not include additional heating 
from transiently heated small grains (see the new radiation 
transfer code of Misselt et al. 2001) which is not important in 
the cooler Classical T~Tauri stars considered in this paper.

\subsection{Energy Sources}

The energy input to the disk is from stellar photons and accretion 
luminosity liberated in the disk.  As discussed in the previous section, 
we fix the disk structure for our radiation transfer simulations.  
Given the disk structure (Eq.~1),
$\alpha$ disk theory determines the accretion rate for a given disk 
mass.  Our parameterization of the disk density and accretion follows 
that presented in the review by Bjorkman (1997), apart from the term 
$\pm\sqrt{R_0/\varpi}$.  The accretion rate and viscosity 
parameter, $\dot M$ and $\alpha_{\rm disk}$, are related to the disk 
parameters by 
\begin{equation}
\dot M = \sqrt{18 \pi^3}\,\alpha_{\rm disk}\, V_c\, \rho_0\, 
h_0^3/R_\star \; ,
\end{equation}
where the critical velocity $V_c=\sqrt{GM_\star/R_\star}$.  The flux 
due to viscous disk accretion, $GM_\star\dot M/2R_\star$, 
is generated throughout the disk midplane region according to 
(Shakura \& Sunyaev 1973, Lynden-Bell \& Pringle 1974)
\begin{equation}
{{dE}\over{dA\, dt}}={{3GM_\star\dot M}\over{4\pi \varpi^3}}
\left [ 1-\sqrt{{R_\star}\over{\varpi}}\right ]\; .
\end{equation}
For low mass disks, the heating due to accretion luminosity 
is negligible and stellar irradiation dominates the disk heating. 
In general we choose $\alpha_{\rm disk}=0.01$ (Hartmann et al. 1998), 
but for the most massive disk we simulate this results in a very large 
accretion luminosity, $L_A>0.8L_\star$.  For this case 
$\alpha_{\rm disk}$ is adjusted so that $L_A<0.2L_\star$ in line with recent 
observational determinations of accretion luminosities in classical T~Tauri 
stars (Hartmann et al. 1998).

\section{Model Results: SEDs, Colors}

The following models use $T_\star=4000$K, $R_\star = 2R_\odot$, 
$R_{\rm disk}= 100$AU, and a distance of 500pc to the system.  

\subsection{SED Evolution with Disk Mass}

Figure 2 shows 
the effect on the SED of changing the disk mass, but keeping the disk 
structure fixed to that of our fiducial model.  The massive optically thick 
disks produce SEDs that resemble that of the 
Kenyon \& Hartmann (1987) flared disk model, aside from differences 
(addition of scattered light to pole-on views and silicate features) 
due to our inclusion of a finite albedo and non-gray opacity.  The most 
massive disk has a very large near-IR excess due to the large accretion 
luminosity present in this model.  
The model SEDs display the characteristic 
features present in other simulations: 
large infrared excess emission, flat spectrum sources at intermediate 
inclinations, and 
double peaked spectra (optical and far-IR peaks) for very large 
inclinations.  At large inclinations, the optical peak is due to scattered 
starlight as the dense disk totally 
obscures the star and disk emission at short wavelengths.  

Reducing the disk mass has the most dramatic 
effect at long wavelengths with the SED rapidly declining with decreasing 
mass.  At short wavelengths the disk remains 
optically thick to stellar photons over a wide range of mass so the 
near-IR excess is not as sensitive to mass.
Figure~4 presents our models again, with the three panels showing 
SEDs for the range of disk masses at a given inclination.  These results show 
similar features to the models of Men'shchikov \& Henning 
(1997, their Fig.~12) who 
presented SEDs for a range of optical depths in a spherical geometry with 
evacuated bipolar cones.  

The {\it SIRTF} sensitivity limits show that at a distance of 500pc 
very low mass disks ($M_{\rm disk}\ga 10^{-6}M_\odot$) are detectable 
out to a wavelength of $70\mu{\rm m}$.  
At $25\mu {\rm m}$ {\it SIRTF} will be sensitive to photospheric 
flux levels at 500pc, thus allowing for the detection of even lower 
mass disks.  Our overall conclusion from these simulations is that 
near-IR excesses detect disks while far-IR excesses can be used 
to study their mass.  

\subsection{Color Evolution with Disk Mass}

Figure~4 shows the variation of colors with disk mass for $i<60^\circ$.  
At long wavelengths we use the simulated flux at $70\mu{\rm m}$ and 
$160\mu{\rm m}$ in forming the color and have not adopted any 
particular color system.  The colors are fairly insensitive to inclination 
for $i<60^\circ$, but see \S3.3 for color-color and color-magnitude diagrams 
that include highly inclined disks.  
For a passive disk with $M_{\rm disk} = 10^{-1}M_\odot$, 
$\Delta (K-L)\approx 0.7$, 
decreasing to a fairly constant $\Delta (K-L) \approx 0.4$ for 
$10^{-7}M_\odot \le M_{\rm disk} \le 10^{-3}M_\odot$.  Larger $K-L$ 
colors arise in massive disks where accretion luminosity is included, 
$\Delta (K-L)\approx 1$ for $M_{\rm disk} = 10^{-1}M_\odot$.  
In our models, less massive disks do not sustain large accretion 
rates (see \S2.3) and the $K-L$ excess is due to reprocessing of 
starlight.  For these disks, $K-L$ is fairly insensitive to mass as the 
disk remains optically thick in the near-IR for masses 
$M_{\rm disk} \ga 10^{-7}M_\odot$.  Therefore ground based near-IR 
observations are capable of detecting very low mass disks.  Again, we 
emphasize that this is the mass of small ($\la 50\mu{\rm m}$) 
particles which dominate the near-IR opacity.

At longer wavelengths the SED decreases with decreasing disk mass and 
this is reflected in the other color indexes remaining relatively flat 
with disk 
mass until the disk becomes optically thin at the waveband under study.  
At the longest 
{\it MIPS} wavelength, $160\mu$m, the $K-160$ color shows a clear 
progression from the most massive to least massive disk.  

\subsection{Color-Color and Color-Magnitude Diagrams}

Figure~5 shows color-color diagrams for our model disks assuming intrinsic 
stellar colors from Kenyon \& Hartmann (1995, Table A5).   
When we include all disk inclinations there are 
degeneracies in the color-mass parameter space, with large viewing angles 
generally leading to redder colors.  This result differs from other 
investigations (e.g., Lada \& Adams 1992; Kenyon, Yi, \& Hartmann 1996; 
Meyer, Calvet, \& Hillenbrand 1997) which assumed the disk emission 
was proportional to $\cos i$, giving the bluest colors for edge-on 
viewing.  More detailed radiation transfer modeling 
shows that a simple $\cos i$ scaling does not give the correct inclination 
dependence for flared disk models.

The effect of 
changing the disk mass is not very strong in the $(K-L)/(K-N)$ diagram 
for $M_{\rm disk}\ga 10^{-6}M_\odot$, 
reflecting that at short wavelengths the disks remain optically thick over 
a wide range of mass.  The disk masses separate out more clearly in the 
$(K-N)/(K-70)$ diagram.  Therefore color-color 
diagrams of large numbers of sources that compare near and far infrared 
colors provide a means of determining the range of disk masses within a 
cluster.  

Figure~5 showed that inclination effects lead to a large spread in the 
location of different disk masses in the color-color diagrams.  This 
leads to degeneracies in the location of different disk masses.  
However, Fig.~6 shows how color-magnitude diagrams 
can help in breaking the mass-inclination degeneracy.
Placing all our disk models on color-magnitude diagrams shows that 
disks viewed at large inclinations occupy the 
lower righthand corners, i.e., highly inclined disks are faint and red.

\section{SED and Color Dependence on Other Disk Properties}

\subsection{Flat Disks}

The previous sections presented SEDs for disks which have the 
flaring parameters that fit HH30~IRS images and SED.  However, other 
investigations indicate a 
variety of disk structures in Classical T~Tauri stars 
(e.g., Kenyon \& Hartmann 1987; Miyake \& Nakagawa 1995; 
Chiang et al. 2001).  
To investigate the effects of flatter disks on SEDs and colors 
we repeated the simulations of Fig.~2 with a reduced scaleheight 
$h_0=0.003R_\star$ giving $h(100{\rm AU}) = 3$AU.  The smaller scaleheight 
results in the disk intercepting less stellar radiation (e.g., Kenyon \& 
Hartmann 1987) and consequently smaller excesses and bluer colors than our 
models in \S~3.  For a given disk mass, the mm flux is unaltered, but the mid 
to far-IR emission is sensitive to the disk structure (Kenyon \& Hartmann 
1987; Chiang et al. 2001).  
The qualitative variation of the SEDs and colors with 
disk mass remains the same.

\subsection{Grain Growth}

In our models the dust size distribution does not change as the 
disk mass decreases.  However, small dust grains may coagulate to form 
larger grains and rocks thereby altering the grain size distribution in 
the disk (e.g., Beckwith, Henning, \& Nakagawa 2000).  
This will lead to changes in the wavelength dependence of 
the opacity.  D'Alessio et al. (2001) investigated grain growth by keeping 
the disk mass constant and increasing the maximum grain size (effectively 
reducing the population of small grains) and changing the slope of the power 
law size distribution.  For a slope $p=3.5$, their SED models show the 
largest effects occur for $\lambda\ga 20\mu$m, while for $p=2.5$ (which 
puts more mass into larger particles) the SEDs are similar to ours in 
which the total disk mass decreases.  
SEDs arising from either a low mass disk or a massive disk with large grains 
may be distinguished with detailed SED modeling.

\subsection{Disk Radius}

Decreasing the disk radius, but keeping the mass constant results in a  
disk that is denser and optically thicker than our models of \S~3.  
A smaller disk effectively removes cool material at large radii that 
provides the bulk of the long wavelength emission, yielding smaller far-IR 
and mm fluxes (e.g., Beckwith et al. 1990).   The other effect of squeezing 
the disk into a smaller 
volume is an increase in the height above the midplane at which stellar 
photons are absorbed.  This gives a geometrically thicker dust disk that 
can intercept more stellar photons, raising the near and mid-IR emission.
The overall 
effects of a very small disk are to increase the wavelength at which the 
disk becomes optically thin yielding larger $K-L$ colors and a slower 
decline in the $K-180$ color with decreasing mass.  The larger 
optical depth for small disks will lower the minimum disk mass that may be 
detected from $K-L$ colors.  Large radius disks give the opposite of 
these effects: less near and mid-IR, and more far-IR and mm emission.

\subsection{Inner Disk Holes and Non-homologous Disk Evolution}

In order to investigate general trends, we have presented a simple model 
for disk evolution of protoplanetary disks in which the mass decreases 
homologously with time.  
Real disks will be more complicated than this, with the disk radius growing, 
changes in the flaring parameters due to dust settling, and the possibility 
of accretion being terminated through the opening up of large inner disk 
holes and gaps, resulting in ``inside-out'' 
evolution.  Current data does not provide much support for the 
existence of 
large holes in protoplanetary disks: if they do not show near-IR excesses, 
mid-IR and mm excesses are usually absent also (Stassun et al.
2001, Haisch, Lada \& Lada 2001a), but there are some exceptions notably
GM~Aur (Koerner, Sargent, \& Beckwith 1993).  
{\it SIRTF} should be able to distinguish between low mass
protoplanetary disks, whose mass has 
evolved homologously (without creating large inner holes),
and disks which have evolved by creating large inner holes  
and thus directly test the inside-out disk clearing scenario.

Debris disks often exhibit large 
inner holes or a ring-like structure (e.g., Koerner et al. 1998; 
Jayawardhana et al. 1998; Schneider et al. 1999). 
However, debris
disks are not likely to be protoplanetary disks that have evolved
large inner holes. Debris disks are sufficiently old that
the dust in them is not remnant protoplanetary material but 
instead is created and continually replenished by collisions
between planetessimals formed previously in the disk.
Moreover, protoplanetary disks which simply evolve from the inside out
by clearing large inner holes (but otherwise maintaining their initial 
structure), will still contain large reservoirs
of material in their outer regions and thus should be considerably 
more massive than debris disks.

\section{Comparison to Observations: Disk Masses in Taurus-Auriga}

Observations at mm and radio wavelengths, where the 
disk is optically thin, provide the best probe of disk mass (e.g., 
Beckwith et al. 1990).   When such data is not available our models 
show that far-IR, and in certain circumstances even mid-IR data,
can provide another means of probing disk mass (if 
the disks are not too massive and optically thick).
For example, Kenyon \& Hartmann (1995) presented a compilation of 
optical through far-IR observations of Taurus-Auriga sources 
and noted a pronounced gap in the $K-N$ distribution between the 
bluest Class~II and the reddest Class~III sources. 
Figure~7 shows the $(K-L)/(K-N)$ color-color diagram 
for the Taurus-Auriga sources 
along with our face-on model colors for various disk masses.  In the 
context of our models, the gap in the $K-N$ distribution corresponds to 
disks with $M_{\rm disk}\la 10^{-6}M_\odot$ and indicates that there
are few circumstellar disks with masses between $10^{-6}$ and $10^{-8}$
$M_\odot$ in the Taurus-Auriga cloud (of course, this disk mass does 
not include rocks and planets).  This, in turn, could be interpreted
to indicate that once the disk mass falls below $\sim 10^{-6} M_\odot$
the timescale for clearing the remaining material in the 
inner ($\la 5-10$AU) disk is very rapid.

\section{Summary}

We have investigated the observational signatures of an evolutionary 
model in which the disk mass decreases homologously, 
but the disk structure and stellar parameters remain constant with time.  
Our main conclusions are:

\begin{itemize}

\item {\bf \it Near-IR observations detect disks.}  Disks remain optically 
thick in the near-IR over a wide range of disk mass resulting in measurable 
$K-L$ excesses for disks down to $M_{\rm disk}\sim10^{-7}M_\odot$.  This 
corresponds to a dust mass of $M_{\rm dust}\sim10^{-9}M_\odot$.

\item {\bf \it Mid-IR observations probe disk structure.}  Observations 
in the $20\mu{\rm m}\la\lambda \la 100\mu{\rm m}$ range are sensitive to 
the disk flaring parameters and are crucial for determining the degree of 
dust settling.

\item {\bf \it Far-IR most sensitive to disk mass.}  The far-IR emission 
decreases rapidly with disk mass resulting in a strong correlation between 
$K-160$ and disk mass.  Our simulations indicate that at a distance of 500pc, 
{\it SIRTF} will be able to detect $M_{\rm disk}\ga 10^{-6}M_\odot$ with 
a 500s exposure at $70\mu{\rm m}$ and even lower masses at $25\mu{\rm m}$.

\item {\bf \it General trends are reproduced for a range of disk 
parameters.} Despite degeneracies in fitting SEDs of particular sources, 
we find that varying the disk parameters within generally accepted 
limits introduces a spread in the color-mass relationship, but does not 
affect the general trends.

\end{itemize}

The degeneracies inherent in fitting disk models to wavelength restricted 
datasets illustrates that it is necessary to have full SED 
coverage to constrain the disk structure and dust size distribution 
(e.g., Men'shchikov \& Henning 1997; 
Chiang et al. 2001; D'Alessio et al. 2001).  
{\it SIRTF}, {\it SMA}, and {\it ALMA} will 
provide detailed SEDs for nearby sources enabling detailed modeling of the 
disk mass and structure.  For studies of stellar clusters, 
such data will not be available and mm observations, 
which are most sensitive to disk mass, cannot as yet achieve the required 
sensitivity and resolution to study low mass disks and/or distant clusters.  
Therefore, we must appeal to statistics and determine average 
properties of sources in different clusters for studying disk evolution.  
{\it SIRTF} will provide broad band colors for hundreds of sources within 
a given cluster, from which we 
can construct median colors and the spread around the median for each 
cluster (e.g., Haisch, Lada, \& Lada 2001b).  From our models it 
is apparent that NIR excesses detect disks, and therefore the absence of 
a $K-L$ excess within a cluster of a given age can determine the maximum 
disk lifetime.  
By comparing the median colors from different clusters with a 
spread in age (as a whole and as functions of position within the cluster), 
we can search for trends as a function of cluster age.  Therefore, the 
far-IR probed by {\it SIRTF} may allow us to determine a timescale for 
disk evolution --- i.e., in our homologous evolutionary model we would be 
able to convert the color-mass correlation to one which tracks disk mass 
in clusters as a function of age, analogous to the recent {\it ISO} analysis 
of debris disks by Spangler et al. (2001).

\acknowledgements

We acknowledge financial support from NASA's Long Term Space Astrophysics 
Research Program, NAG5~6039 (KW), NAG5~8412 (BW), NAG5~7993 (MW), 
NAG5~3248 (JEB);
the National Science Foundation, AST~9909966 (BW and KW), AST~9819928 (JEB), 
and a PPARC Advanced Fellowship (KW).

\newpage

\begin{figure}[t]
\centerline{\plotfiddle{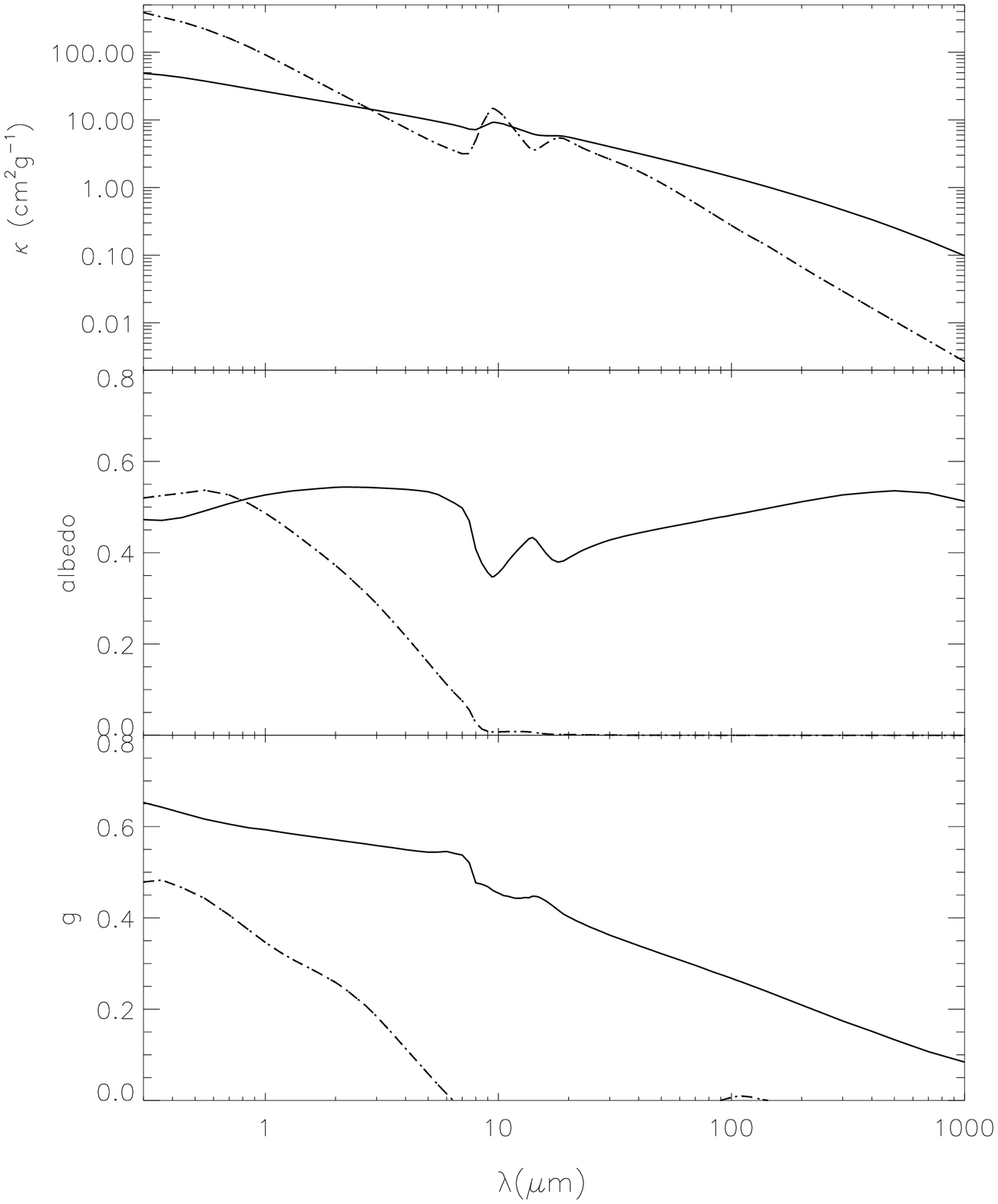}{6in}{0}{65}{65}{-460}{0}}
\caption{Dust parameters for a grain size distribution that fits the 
scattered light images and SED of HH30~IRS (solid line).  The dashed lines 
show ISM grain parameters (Kim, Martin, \& Hendry 1994).  
The three panels show total opacity (upper), albedo 
(middle), and cosine asymmetry parameter (lower).}
\end{figure}

\begin{figure}[t]
\centerline{\plotfiddle{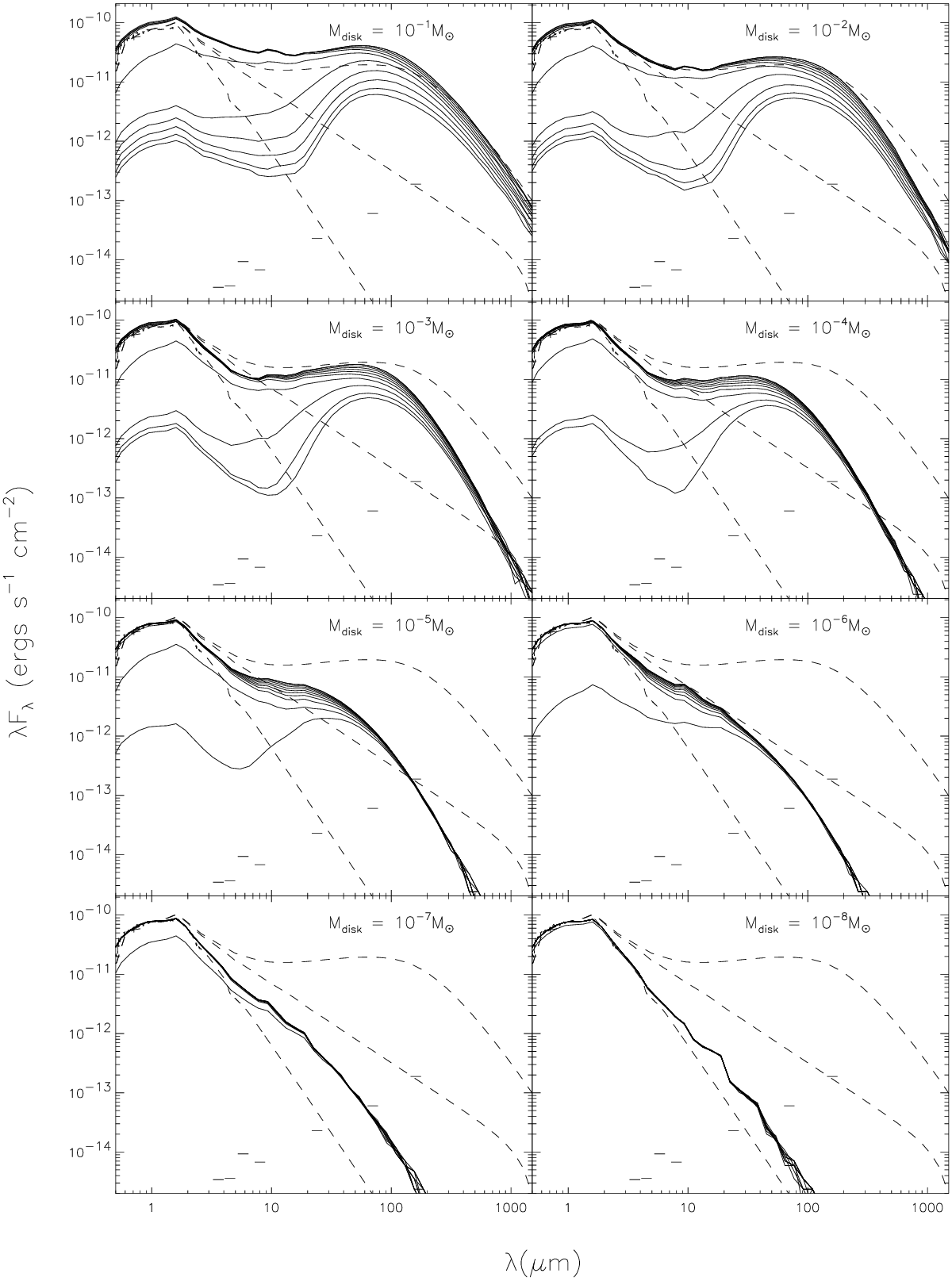}{6in}{0}{65}{65}{-420}{30}}
\caption{Evolution of SED with disk mass.  Each panel shows model SED at 
ten viewing angles (evenly spaced in $\cos i$), ranging from $i=87^\circ$ 
(lowest curve) to $i=13^\circ$ (upper curve).  Also shown are the input 
stellar spectrum and optically thick flat and flared reprocessing disk 
models.  The 
star is assumed to be at a distance of 500pc.  The short horizontal lines 
are $5\sigma$ {\it SIRTF} sensitivity limits for a 500~s exposure 
(http://sirtf.jpl.nasa.gov/SSC/B\_Observing/SSC\_B2.html).}
\end{figure}

\begin{figure}[t]
\centerline{\plotfiddle{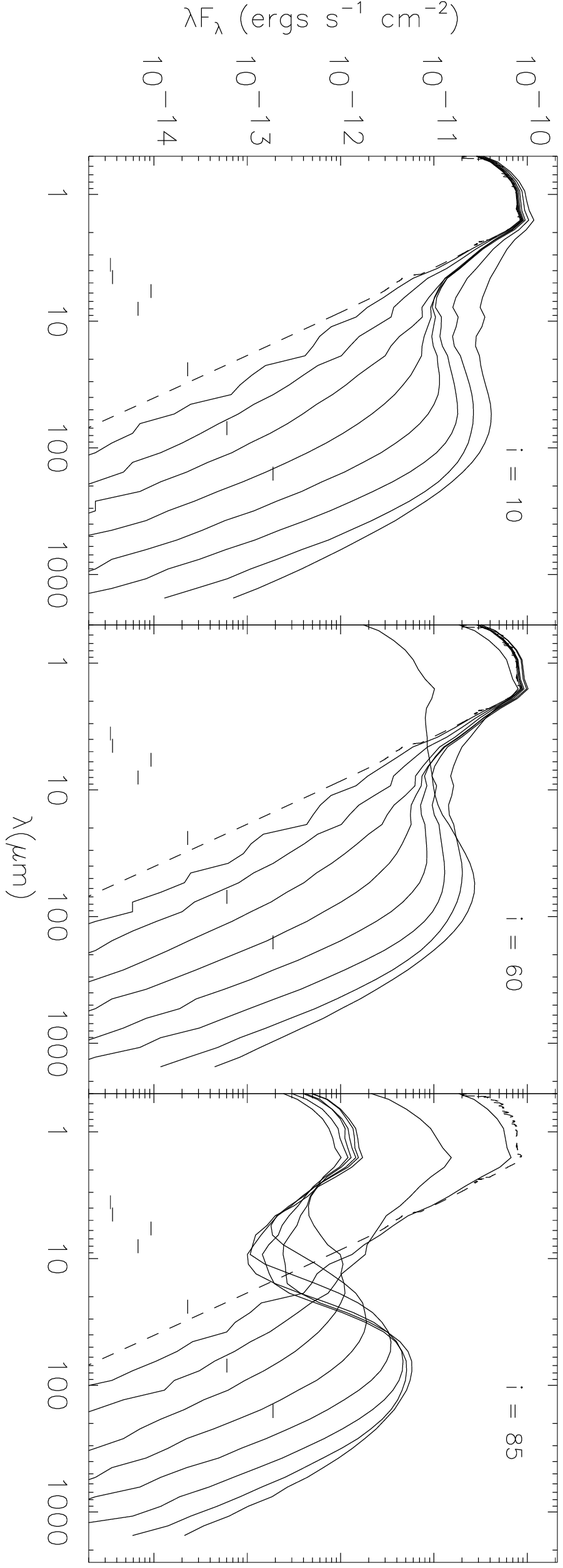}{6in}{90}{65}{65}{0}{0}}
\caption{Evolution of SED with disk mass. Each panel shows model SEDs for 
a fixed viewing angle for the range of disk masses, 
$M_{\rm disk}=10^{-8}M_\odot$ (lowest curve) to 
$M_{\rm disk}=10^{-1}M_\odot$ (upper curve).  Input spectrum and {\it SIRTF} 
sensitivities are as in Fig.~3.}
\end{figure}

\begin{figure}[t]
\centerline{\plotfiddle{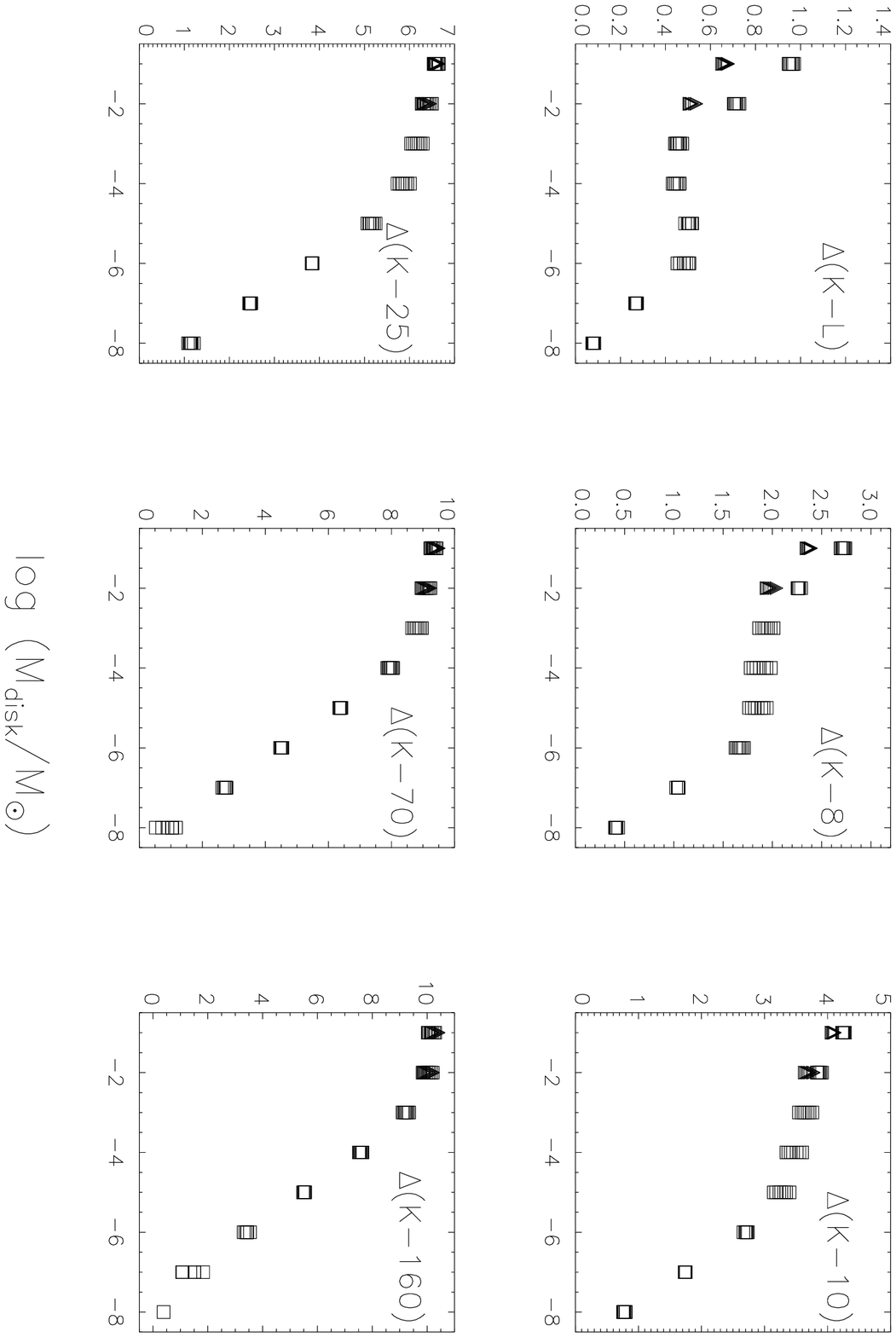}{6in}{90}{65}{65}{0}{0}}
\caption{Colors as a function of disk mass for inclinations $i<60^\circ$.  
Triangles are passive disks, squares include accretion luminosity.  
This bottom panels show that long wavelengths are more sensitive to disk 
mass.}
\end{figure}

\begin{figure}[t]
\centerline{\plotfiddle{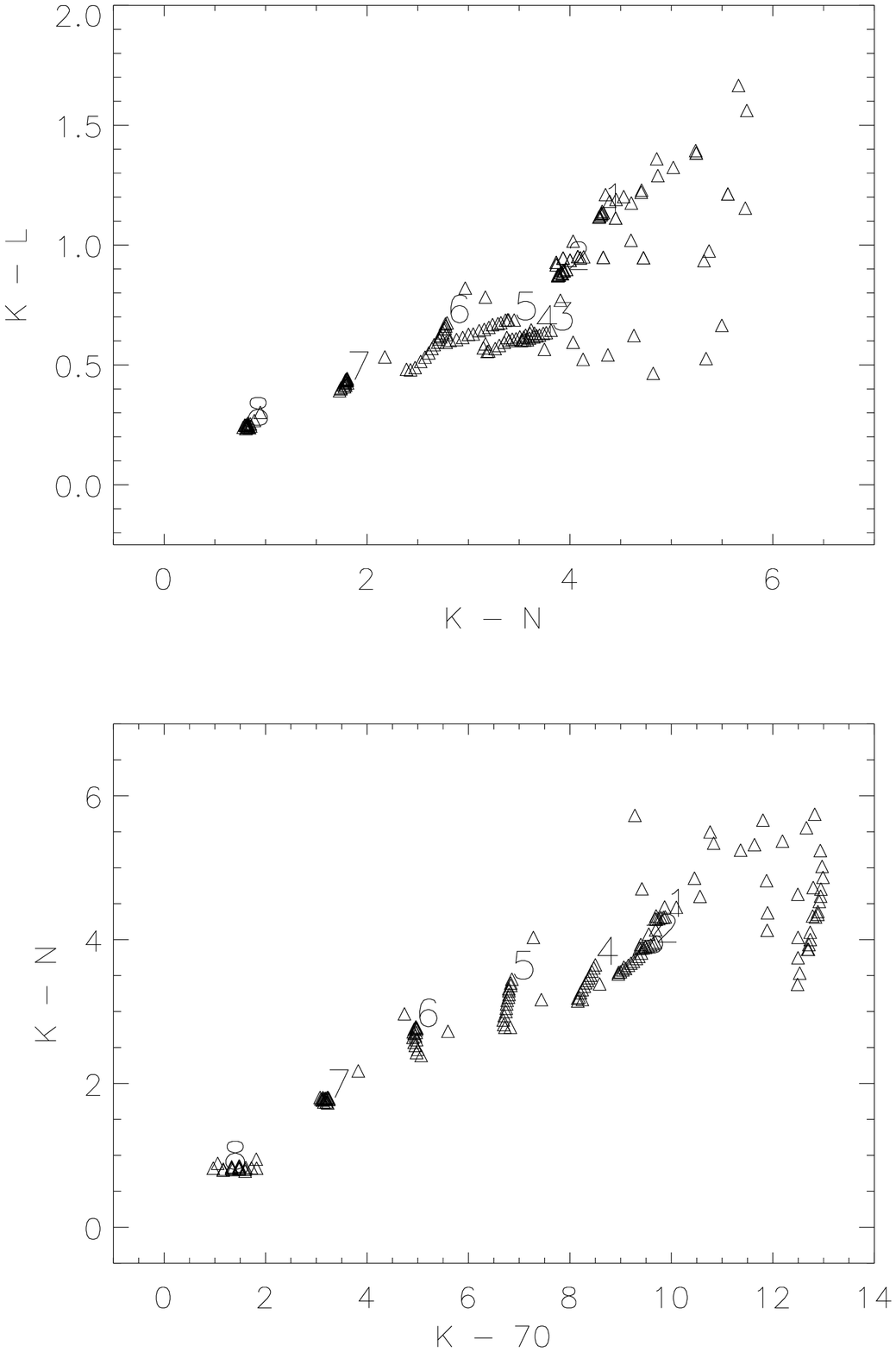}{6in}{0}{65}{65}{-420}{0}}
\caption{Color-color diagrams for our models.  For each model we show twenty 
inclinations, evenly spaced in $\cos i$.  
The numbers are the locations of different mass disks viewed pole on, 
$10^{-1}M_\odot = 1$, $10^{-2}M_\odot = 2$, $10^{-3}M_\odot = 3$, etc.  
The very red colors are the highly inclined disks.  As with Fig.~4, 
the bottom panel shows that long wavelength colors can distinguish a wide 
range of disk masses.}
\end{figure}

\begin{figure}[t]
\centerline{\plotfiddle{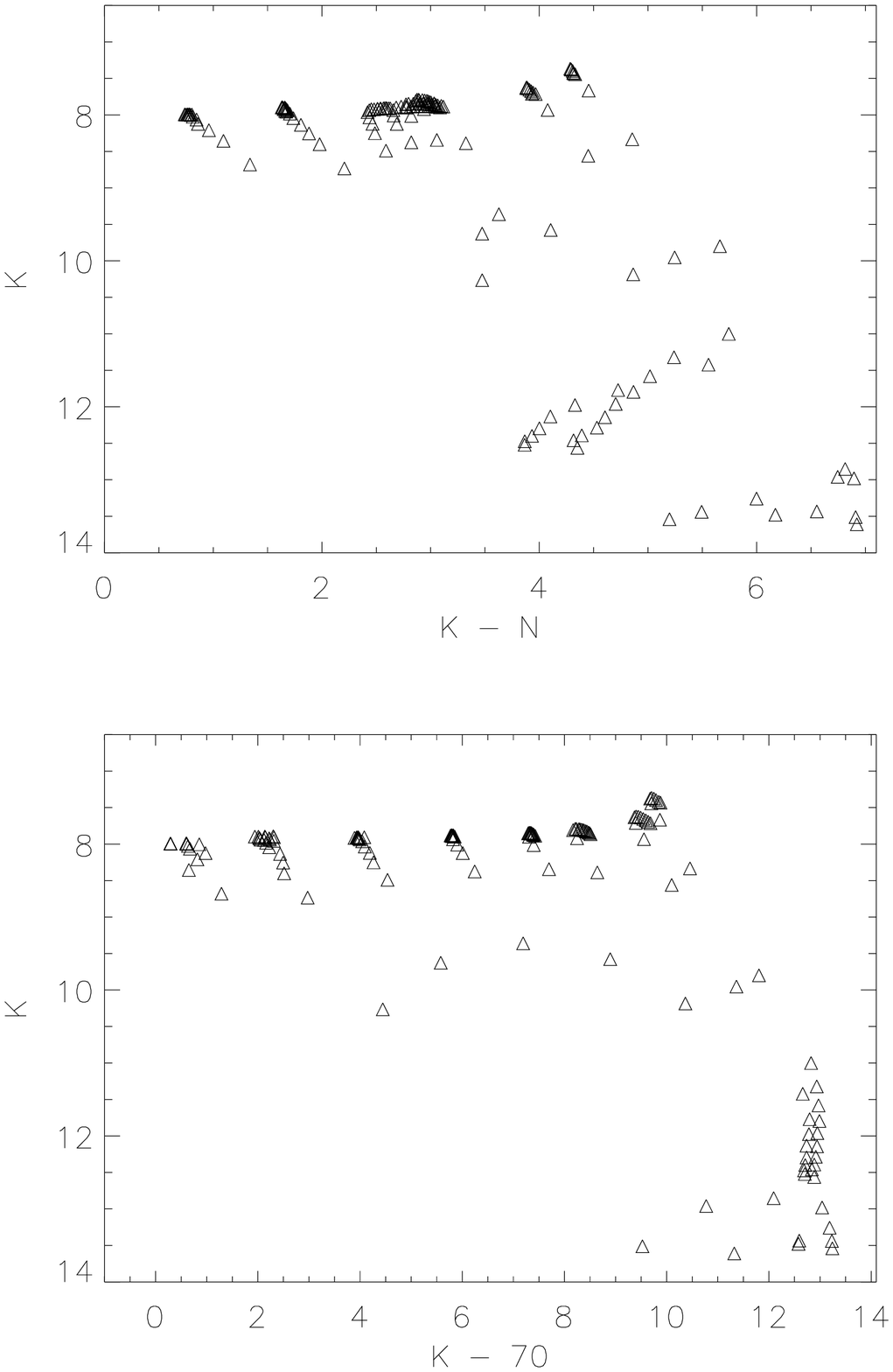}{6in}{0}{65}{65}{-420}{0}}
\caption{Color-magnitude diagrams for our models assuming a 
distance of 500pc.  
The edge-on disk sources are very faint and red and occupy the lower 
right corners of the color-magnitude diagrams. }
\end{figure}

\begin{figure}[t]
\centerline{\plotfiddle{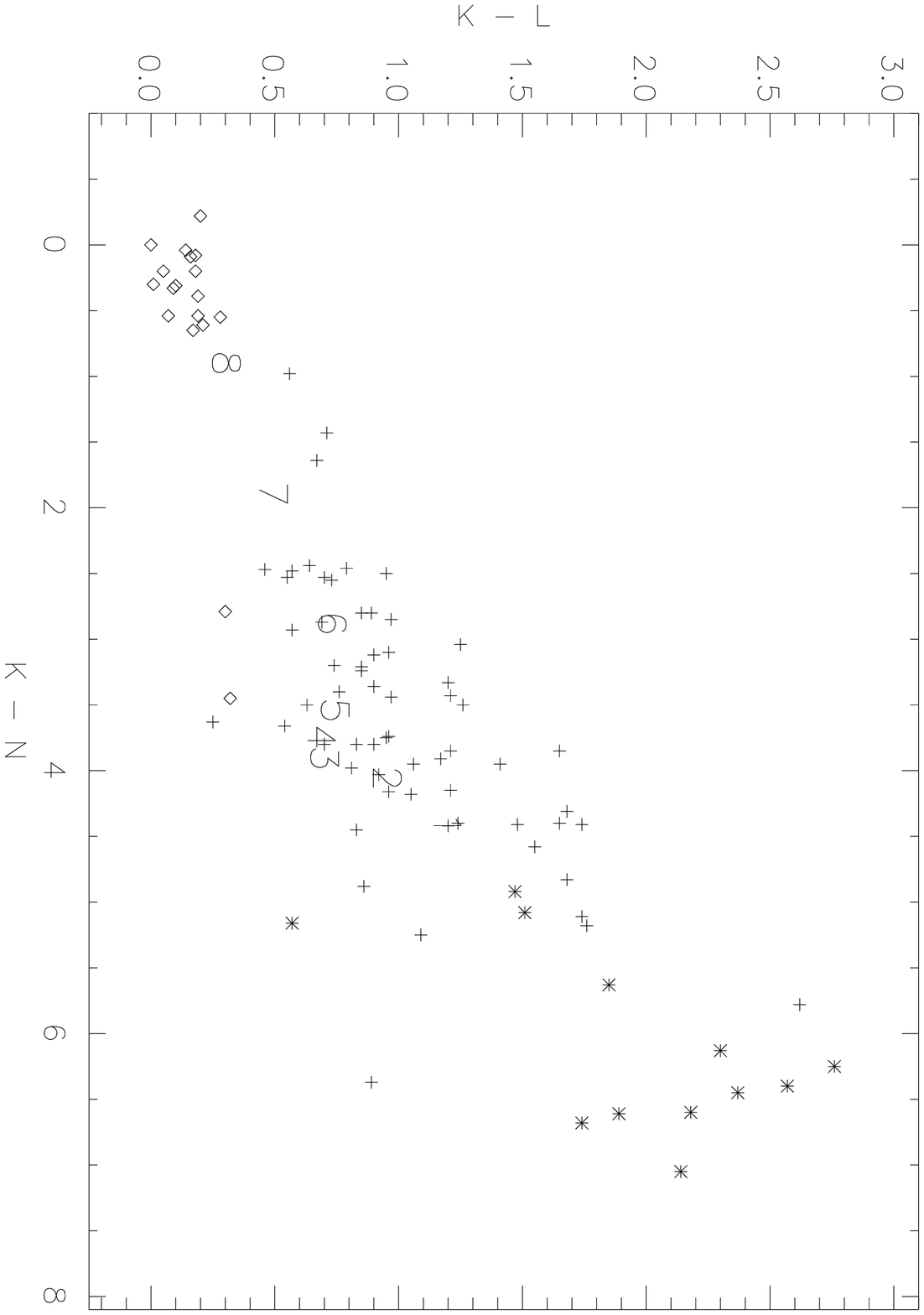}{6in}{90}{65}{65}{-50}{0}}
\caption{Color-color diagrams for Taurus-Auriga sources Class~III 
(diamonds), Class~II (crosses), Class~I (stars).  The data are taken from 
the Kenyon \& Hartmann (1995) compilation.  The numbers are our 
models for different mass disks viewed pole-on as in Fig.~5.  The gap 
in the $K-N$ distribution between the Class~II and Class~III 
sources is filled in by our models having $M_{\rm disk}\la 10^{-6}M_\odot$, 
which may imply that the timescale to clear $M_{\rm disk}\la 10^{-6}M_\odot$ 
is very rapid.}
\end{figure}


\begin{thebibliography}{}

\bibitem[]{}
Adams, F.C., Emerson, J.P., \& Fuller, G.A. 1990, ApJ, 357, 606

\bibitem[]{}
Adams, F.C., Lada, C.J., \& Shu, F.H. 1987, ApJ, 308, 788

\bibitem[]{}
Adams, F.C., \& Shu, F.H. 1986, ApJ, 308, 836

\bibitem[]{}
Backman, D.E., \& Paresce, F. 1993, in Protostars and Planets III, 
ed. E.H. Levy \& J.I. Lunine (Tuscon: Univ. Arizona Press), 1253

\bibitem[]{}
Beckwith, S. V. W., Henning, T., \& Nakagawa, Y. in 
``Protostars and Planets IV,'' 
eds Mannings, V., Boss, A.P.,\&  Russell, S. S., 
(Tucson: University of Arizona Press), 533

\bibitem[]{}
Beckwith, S.V.W., \& Sargent, A.I. 1993, in Protostars \& Planets III, 
ed. E.H. Levy \& J.I. Lunine (Tucson: Univ. Arizona), 543

\bibitem[]{}
Beckwith, S.V.W., \& Sargent, A.I. 1991, ApJ, 381, 205

\bibitem[]{}
Beckwith, S.V.W., Sargent, A.I., Chini, R.S., \& Gusten, R. 1990, 
AJ, 99, 924

\bibitem[]{}
Bell, K.R. 1999, ApJ, 526, 411

\bibitem[]{}
Bell, K.R., Cassen, P.M., Klahr, H.H., \& Henning, T. 1997, ApJ, 486, 372

\bibitem[]{}
Bjorkman, J.E., \& Wood, K. 2001, ApJ, 554, 615

\bibitem[]{}
Bjorkman, J.E. 1997, in ``Stellar Atmospheres : Theory and Observations,''
eds J.P. De Greve, R. Blomme, \& H. Hensberge (Springer)

\bibitem[]{}
Boss, A.P., \& Yorke, H.W. 1996, ApJ, 469, 366

\bibitem[]{}
Burrows, C.J., et al. 1996, ApJ, 473, 437

\bibitem[]{}
Calvet, N., Canto, J., Binette, L., \& Raga, A.C. 1992, Rev. Mex., 24, 81

\bibitem[]{}
Chiang, E.I., Joung, M.K., Creech-Eakman, M.J., Qi, C., Kessler, J.E., 
Blake, G.A., \& van Dishoeck, E.F. 2001, ApJ, 547, 1077

\bibitem[]{}
Chiang, E.I., \& Goldreich, P. 1999, ApJ, 519, 279

\bibitem[]{}
Chiang, E.I., \& Goldreich, P. 1997, ApJ, 490, 368

\bibitem[]{}
Cohen, M., \& Kuhi, L.V. 1979, ApJS, 41, 743

\bibitem[]{}
Cotera, A.S., et al. 2001, ApJ, 556, 958

\bibitem[]{}
D'Alessio, P., Calvet, N., \& Hartmann, L. 2001, ApJ, 553, 321

\bibitem[]{}
D'Alessio, P., Calvet, N., Hartmann, L., Lizano, S., \& Canto, J. 
1999, ApJ, 527, 893

\bibitem[]{}
D'Alessio, P., Canto, J., Calvet, N., \& Lizano, S. 1998, ApJ, 500, 411

\bibitem[]{}
Dutrey, A., Guilloteau, S., Duvert, G., Prato, L., Simon, M., 
Schuster, K., \& Menard, F. 1996, A\&A, 309, 493

\bibitem[]{}
Efstathiou, A., \& Rowan-Robinson, M. 1991, MNRAS, 252, 528

\bibitem[]{}
Grady, C.A., et al. 2000, ApJ, 544, 895

\bibitem[]{}
Haisch, K.E., Lada, E.A., \& Lada, C.J. 2001a, AJ, 121, 2065

\bibitem[]{}
Haisch, K.E., Lada, E.A., \& Lada, C.J. 2001b, ApJ, 553, L153

\bibitem[]{}
Haisch, K.E., Lada, E.A., \& Lada, C.J. 2000, AJ, 120, 1396

\bibitem[]{}
Hartmann, L., Calvet, N., Gullbring, E., \& D'Alessio, P. 1998, ApJ, 
495, 385

\bibitem[]{}
Henyey, L.C., \& Greenstein, J.L. 1941, ApJ, 93, 70

\bibitem[]{}
Hillenbrand, L.A., Strom, S.E., Vrba, F.J., \& Keene, J. 1992, ApJ, 
397, 613

\bibitem[]{}
Jayawardhana, R., Fisher, S., Hartmann, L., Telesco, C., Pina, R., 
\& Razio, G. 1998, ApJ, 503, L79

\bibitem[]{}
Kenyon, S.J., Yi, I., \& Hartmann, L. 1996, ApJ, 462, 439

\bibitem[]{}
Kenyon, S.J., \& Hartmann, L. 1995, ApJS, 101, 117

\bibitem[]{}
Kenyon, S.J., \& Hartmann, L. 1987, ApJ, 323, 714

\bibitem[]{}
Kim, S.H., Martin, P.G., \& Hendry, P.D. 1994, ApJ, 422, 164

\bibitem[]{}
Koerner, D.W., Ressler, M.E., Werner, M.W., \& Backman, D.E. 1998, ApJ, 
505, L83

\bibitem[]{}
Koerner, D.W., \& Sargent, A.I. 1995, AJ, 109, 2138

\bibitem[]{}
Koerner, D.W., Sargent, A.I., \& Beckwith, S.V.W. 1993, ApJ, 408, L93

\bibitem[]{}
Koresko, C.D. 1998, ApJ, 507, L145

\bibitem[]{}
Krist, J.E., Stapelfeldt, K.R., Menard, F., Padgett, D.L., \& Burrows, C.J. 
2000, ApJ, 538, 793

\bibitem[]{}
Lada, C.J., \& Adams, F.C. 1992, ApJ, 393, 728

\bibitem[]{}
Lynden-Bell, D., \& Pringle, J.E. 1974, MNRAS, 168, 603

\bibitem[]{}
Mendoza, E.E. 1968, ApJ, 151, 977

\bibitem[]{}
Men'shchikov, A.B., \& Henning, T. 1997, A\&A, 318, 879

\bibitem[]{}
Meyer, M.R., Calvet, N., \& Hillenbrand, L.A. 1997, AJ, 114, 288

\bibitem[]{}
Miyake, K., \& Nakagawa, Y. 1995, ApJ, 441, 361

\bibitem[]{}
O'Dell, C.R., Wen, Z., \& Hu, X. 1993, ApJ, 410, 696

\bibitem[]{}
Padgett, D.L., Brandner, W., Stapelfeldt, K.R., Strom, S.E., Terebey, S., 
\& Koerner, D.W. 1999, AJ, 117, 1490

\bibitem[]{}
Roddier, C., Roddier, F., Northcott, M.J., Graves, J.E., Jim, K. 1996, 
ApJ, 463, 326

\bibitem[]{}
Rucinski, S.M. 1985, AJ, 90, 2321

\bibitem[]{}
Rydgren, A.E., Strom, S.E., \& Strom, K.M. 1976, ApJS, 30, 307

\bibitem[]{}
Shakura, N.I., \& Sunyaev, R.A. 1973, A\&A, 24, 337

\bibitem[]{}
Schneider, G., et al. 1999, ApJ, 513, L127

\bibitem[]{}
Shu, F.H., Adams, F.C., \& Lizano, S. 1987, ARA\&A, 25, 23

\bibitem[]{}
Sonnhalter, C., Preibisch, T., \& Yorke, H.W. 1995, A\&A, 299, 144

\bibitem[]{}
Spangler, C., Sargent, A.I., Silverstone, M.D., Becklin, E.E., 
\& Zuckerman, B. 2001, ApJ, in press, astro-ph/0103185

\bibitem[]{} 
Stapelfeldt, K. R., \& Moneti, A.  1999 in ``The Universe as Seen by ISO'', 
Eds. P. Cox \& M. F. Kessler, ESA-SP 427, 521

\bibitem[]{} 
Stapelfeldt, K. R., Krist, J.E., Menard, F., Bouvier, J., Padgett, D.L., 
\& Burrows, C.J. 1998, ApJ, 502, L65

\bibitem[]{}
Stassun, K.G., Mathieu, R.D., Vrba, F.J., Mazeh, T., \& Henden, A. 2001, 
AJ, 121, 1003

\bibitem[]{}
Thi, W.F., et al. 2001, Nature, 409, 60

\bibitem[]{}
Wilner, D., \& Lay, O. 2000, in ``Protostars and Planets IV,'' 
eds Mannings, V., Boss, A.P.,\&  Russell, S. S., 
(Tucson: University of Arizona Press), p. 509

\bibitem[]{}
Wolf, S., Henning, T., \& Stecklum, B. 1999, A\&A, 349, 839

\bibitem[]{}
Wood, K., Wolff, M.J., Bjorkman, J.E., \& Whitney, B.A. 2001, ApJ, 
in press, astro-ph/0109048

\end{thebibliography}
\end{document}